\documentclass{cernrep} 
\usepackage{texnames}
\usepackage[T1]{fontenc}
\usepackage[bookmarks, colorlinks=true, linktoc=page, linkcolor=black, citecolor=black, urlcolor=blue]{hyperref}
\usepackage{fancyhdr}
\fancyhfoffset{4 mm}
\fancypagestyle{ARTTITLE}{%
\fancyhf{} 
\lhead{\footnotesize{Proceedings of the 2019 CERN--Accelerator--School course on
\it{ High Gradient Wakefield Accelerators}, Sesimbra, (Portugal)}}
\lfoot{Available online at \url{https://cas.web.cern.ch/previous-schools}}
\rfoot{\thepage\hspace*{3mm}}

}
\begin{document}
\title{Particle Beam Diagnostics}
 
\author {Barbara Marchetti}

\institute{DESY, Hamburg, Germany}

\begin{abstract}
This lecture gives an overview about beam diagnostics techniques for the characterization of electron bunches obtained with plasma accelerators. Due to the limited space, the lecture does not aim to go into the specifics of the single measurements but rather to offer an overview of the present techniques (including ongoing developments) highlighting advantages and disadvantages of each of them. The linked references provide on the other side a more in depth discussion about the technical aspects. 
\end{abstract}


\maketitle 
\thispagestyle{ARTTITLE}

\section{Introduction to Particle Beam Diagnostics}
Diagnostics tools are crucial components of particle accelerators. On the one hand they allow to control the particle beam and on the other hand they provide detailed information concerning the stability and the quality of the particle beam delivered at the user-area.\\
Controlling the particle beam means to be able to transport the electrons, tune the working point, feedback the measured data in order to prevent unwanted drifts of the machine settings (caused e.g. by temperature variations), measure beam losses and prevent the electrons from damaging sensitive parts of the accelerator. 
In this lecture we will focus on measurements for particle beam characterization aimed at gaining information concerning the quality and the reproducibility of the electron bunch. The typical quantities to be characterized are:
\begin{itemize}
\item The charge of the electron beam;
\item the transverse emittance and the distribution of the transverse phase space of the beam;
\item the longitudinal phase space of the beam and in particular the beam energy spread and the bunch length.
\end{itemize}
Those quantities are sufficient to characterize the quality of the electron bunch. They indeed allow to calculate the most important figure of merit commonly related to particle beams: the beam brightness.

\subsection{Beam Brightness}
The concept of beam brightness has been introduced in 1939 by von Borries and Ruska (Nobel prize in Physics for the invention of the Electron Microscope). It was initially defined as the current per area $\Delta a$ normal to the beam and per element of solid angle $\Delta \Omega$ along the microscope column. This quantity was found empirically constant along the microscope column and was important for quantifying the ability of the electrons to be transversely focused or collimated.\\
In the accelerator field we are familiar with the so called 5D normalized Brightness, defined as:
\begin{equation}\label{eqbrightness}
B_{n5D}=\frac{2I}{\pi^2 \epsilon_{nx} \epsilon_{ny}} [A/(m rad)^2].
\end{equation}
Compared to the original definition, in Eq.\ref{eqbrightness} the geometrical emittance term has been replaced by the normalized emittance to account for relativistic effects. More details about the derivation of this expression can be found in \cite{brightness}.

\section{Characterization of the traverse phase space of the beam}
\subsection{Transverse Emittance}
The main quantity that characterizes the quality of the phase space of the beam is called \emph{phase space emittance}.\\ 
The derivation of the concept of phase space emittance can be found in beam-dynamics-literature, e.g. in references \cite{buon}, \cite{barletta}, \cite{wiedemann}. Here we will try to synthesize the main concepts which are preparatory to the discussion about beam diagnostics.\\
The evolution of a particle beam along a beamline is defined by the evolution of the position of each particle in a six-dimensional-space having coordinates $(x,p_x,y,p_y,z,p_z)$. In the latter expression we have used the particle position coordinates $x,y,z$ and particle conjugate momenta $p_i$.\\
In presence of only conservative forces and if the motion of the charged particles in the horizontal, vertical and longitudinal planes are uncoupled, the phase space emittance in one of the two-dimensional transverse planes ($(x,p_x)$ and $(y,p_y)$) is conserved. In this case the beam emittance on each plane is defined as the 2D-volume occupied by the particles in the two-dimensional transverse plane.\\
Practically it is more convenient to give a statistical definition of the emittance (called \emph{statistical emittance} or \emph{RMS emittance}). The latter quantity is conserved only under the action of linear forces.\\
Moreover in most of the experiments the phase space emittance is replaced by the so-called \emph{trace space emittance} (also called \emph{geometrical emittance}), which is defined in the planes $(x,x \prime)$ and $(y,y \prime)$, where $x \prime$ and $y \prime$ are the transverse angles with respect to the ideal beam trajectory. The geometrical emittance is conserved in a transport line but not in presence of acceleration.\\
In conventional accelerator field, where beams typically have a small energy spread, the idea of invariant emittance can still be used if the emittance is scaled accordingly to the beam energy. That is why the concept of \emph{normalized tranverse emittance} has been introduced. If the beam energy spread is small, the normalized emittance is conserved in a transport line and during acceleration.\\
In plasma acceleration field anyway the assumption that the beam energy spread is small is typically not valid, therefore the normalized emittance is not conserved, as pointed out in \cite{migliorati} and \cite{li1}.\\
This consideration is of utmost importance for the design of a diagnostics-line for the characterization of the transverse properties of the beam.

\subsection{Challenges of Diagnostics for Plasma Beams}
Besides the usually higher energy spread, beams generated from plasmas have also other characteristics making their diagnose especially challenging.\\
At the present state of the art, they are typically still less stable in energy and position than the beams from conventional accelerators, therefore single shot measurements are highly wished.\\
The beam can have down to sub-fs bunch duration, sub-micrometer transverse emittance and charge down to sub-pC range \cite{overview_diagnostics}. Much research effort is therefore being invested in the testing of novel diagnostics method which could fulfill the needs of this community.

\section{Overview of Techniques for Transverse Emittance Measurements}
In the following paragraph we will compare different techniques for measuring the geometrical RMS transverse emittance of the beam, which on the x-plane is defined as:
\begin{equation}\label{eq_brightness}
\epsilon_{x}=\sqrt{\langle x^2 \rangle \langle ( x^{\prime 2} )\rangle - \langle x x^{\prime} \rangle ^2}.
\end{equation}
This emittance can be calculated by measuring the second order moments of the beam. If the energy spread of the beam is non-negligible, it will be necessary to measure it to calculate the normalized transverse RMS beam emittance\cite{migliorati}. We will deal with energy spread measurements in the next section of this paper.\\
The second order moments of the beam can be measured in two different ways:
\begin{itemize}
\item by looking at the position and divergence of a beam-let (i.e. a sub-portion of the beam) 'surviving' to the interception with a slit, a pepperpot or a grid;
\item by looking at the evolution of the beam envelope in a transport line with fixed or variable element-strengths.
\end{itemize} 
 
\subsection{Slit-Scan, Pepperpot, Transmission Electron Microscopy (TEM) Grid}

\begin{figure}[ht]
\begin{center}
\includegraphics[width=14cm]{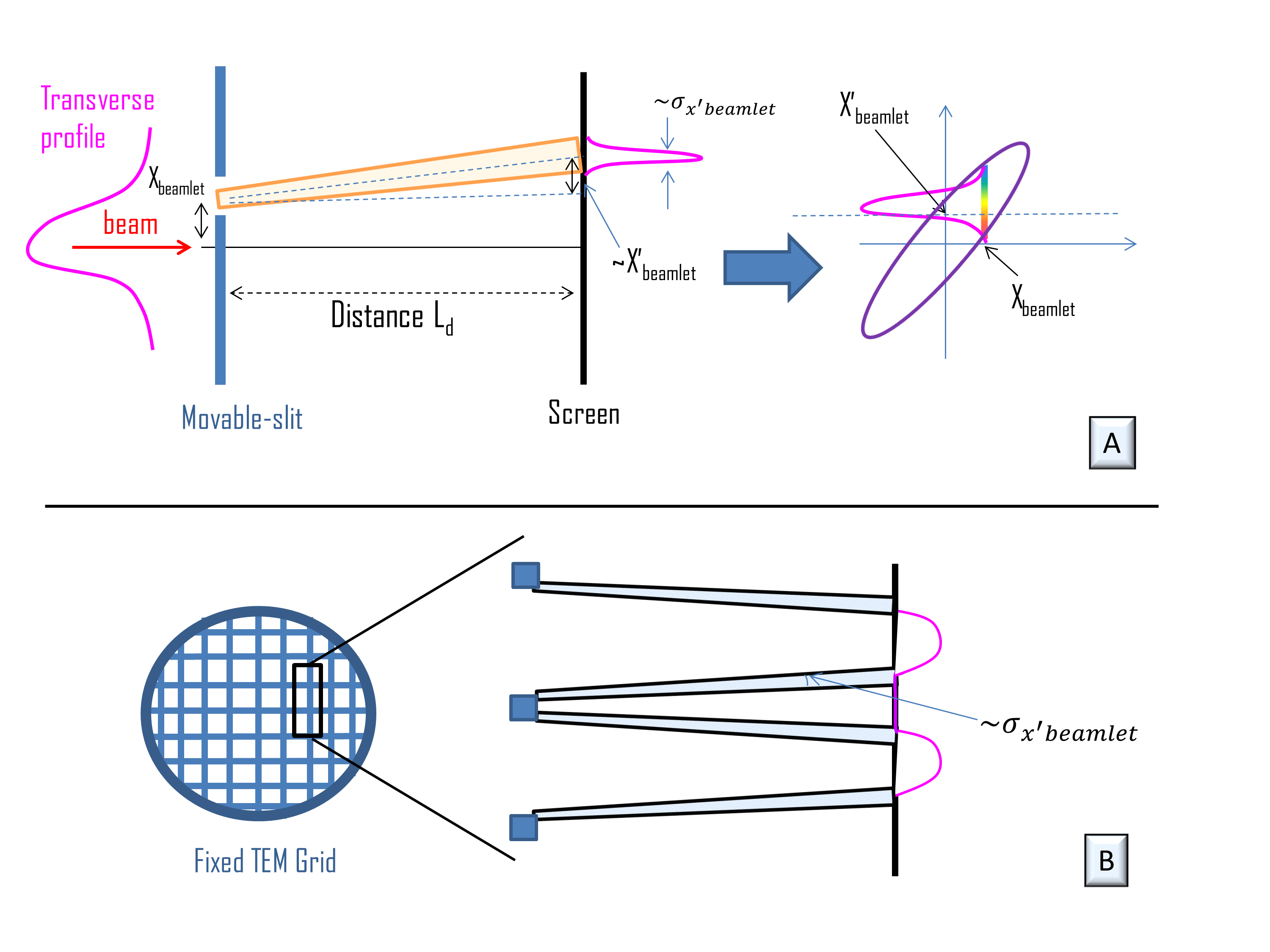}
\caption{Sketch representing the working principle of the slit-scan method (A) and TEM grid method (B) for measuring the beam emittance. }
\label{slits}
\end{center}
\end{figure}

\subsubsection{The Slit Method}
The typical layout for an emittance measurement using a slit-scan is shown in Fig. \ref{slits} A. When the beam impinges on the slit, the slit mask converts a space charge dominated electron beam into an emittance dominated beamlet. The center of the beamlet in X is defined by the position of the slit, which is varied during the scan. From the width and the position of the slit-image on the measurement screen the mean-beamlet-angle and the mean-beamlet-divergence can be computed as explained in \cite{slit_theory1}. By moving the slit across the complete beam distribution in x, the profiles in $x^{\prime}$ of the beam-slices can be reconstructed, as conceptually shown in Fig. \ref{slits} A. Some experimental examples of such measurements can be found e.g. in \cite{slit_theory2} and \cite{pitz}. By looking at the layout it is clear that good resolution can be achieved only if the size of the beamlet at the measurement screen is much bigger than slit-width.

\subsubsection{The Pepperpot Method}
In the Pepperpot technique the movable slit is replaced by a fixed mask with multiple holes having a defined periodic pattern \cite{pepperpot}. The second order moments of the beam distribution can be retrieved in both vertical and horizontal planes by a single-shot measurement using the same formulas described in \cite{slit_theory1}. Contrary to the slit-scan case, in this measurement only a subset of the particle in the beam are analyzed, it is therefore needed to assume that the beam under analysis is homogeneous and the subset of measured particles well represents the complete particle distribution. Also the dimensions of the pepperpot have to be adapted to the particle beam in order to prevent beamlets from overlapping and guarantee sufficient sampling of the beam.  

\subsubsection{The TEM grid Method}
For the characteration of bunches having less than few pico-Coulomb charge it might be difficult to have an acceptable signal to noise ratio using the slit and pepperpot methods. Most of the beam charge is indeed blocked at the slit/pepperpot location and the signal to be analyzed is extremely low. For this reason recently a new method has been proposed that replaces the pepperpot with a TEM grid \cite{TEM1}. In the TEM grid method the information concerning the divergence-spread in the beamlet, called $\sigma_{x^{\prime}}$ in Fig. \ref{slits} B is encoded in the shadow of the edge of the grid-bar and can be retrieved by fitting the distribution as described in \cite{TEM1}. It has to be pointed out that this method works only for low-density, emittance-dominated beams since, contrary to the slit and pepperpot methods, most of the beam charge is not absorbed nor scattered at the grid location.

\subsubsection{Limitations and Recent Developments for Plasma Beams}
Because of its single-shot characteristic, the pepperpot is expecially appealing as an emittance diagnostics method for beams from plasma.\\
Anyway beams from a plasma-injector have typically higher energy and divergence if compared with typical beams characterized at the exit of an RF-injector. In \cite{cianchi} it has been shown that the undersampling error dominates the reconstruction of the transverse trace-space of a beam from plasma.\\
Also there have been recent experiments in the direction of extending the pepperpot method to higher energy ranges. Performing pepperpot at high energies is difficult mainly because a thicker slit is needed. The increased thickness of the pepperpot has important consequences such as the modification of the acceptance of the mask \cite{delerue}. In \cite{thomas} it has been experimentally demonstrated for a beam with 3GeV energy that, if the beam waist is located far enough from the pepperpot, an emittance measurement providing reasonable accuracy is still possible.

\subsection{Quadrupole/Solenoid Scan and Multi-Screen Methods}
The methods described in this section rely on the analysis of the beam envelope evolution along a known transport line including magnets, which might have fixed or variable strength.\\
The analysis of the data is based on the linear matrix transformation approach to represent the transport of the second order moments of the beam from a point $s_0$ at the entrance of the line to the $i^{th}$-measurement-screen located at the positions $s_i$. A detailed overview of the linear beam transport theory and matrix formalism in linear beam dynamics can be found in many references, such as \cite{wiedemann}, \cite{rossbach} and \cite{minty}. In the following we will make use of the equations:
\begin{equation}
\sigma_{s_0}=
\left[ \begin{array}[pos]{ccc} \label{sigma1}
	\langle x^2 \rangle & \langle xx^{\prime} \rangle \\
	\langle xx^{\prime} \rangle & \langle x^{\prime 2} \rangle
\end{array} \right]_{s0}=
\left[ \begin{array}[pos]{ccc}
	\sigma_{11,s_0} & \sigma_{12,s_0} \\
	\sigma_{12,s_0} & \sigma_{22,s_0}
\end{array} \right];
\end{equation}

\begin{equation} \label{sigma2}
\sigma_{s_1}=M \sigma_{s_0} M^T. 
\end{equation}

Where M is the linear transport matrix of the lattice from the longitudinal position $s_0$ to position $s_1$ along te diagnostics line.\\

\begin{figure}[ht]
\begin{center}
\includegraphics[width=14cm]{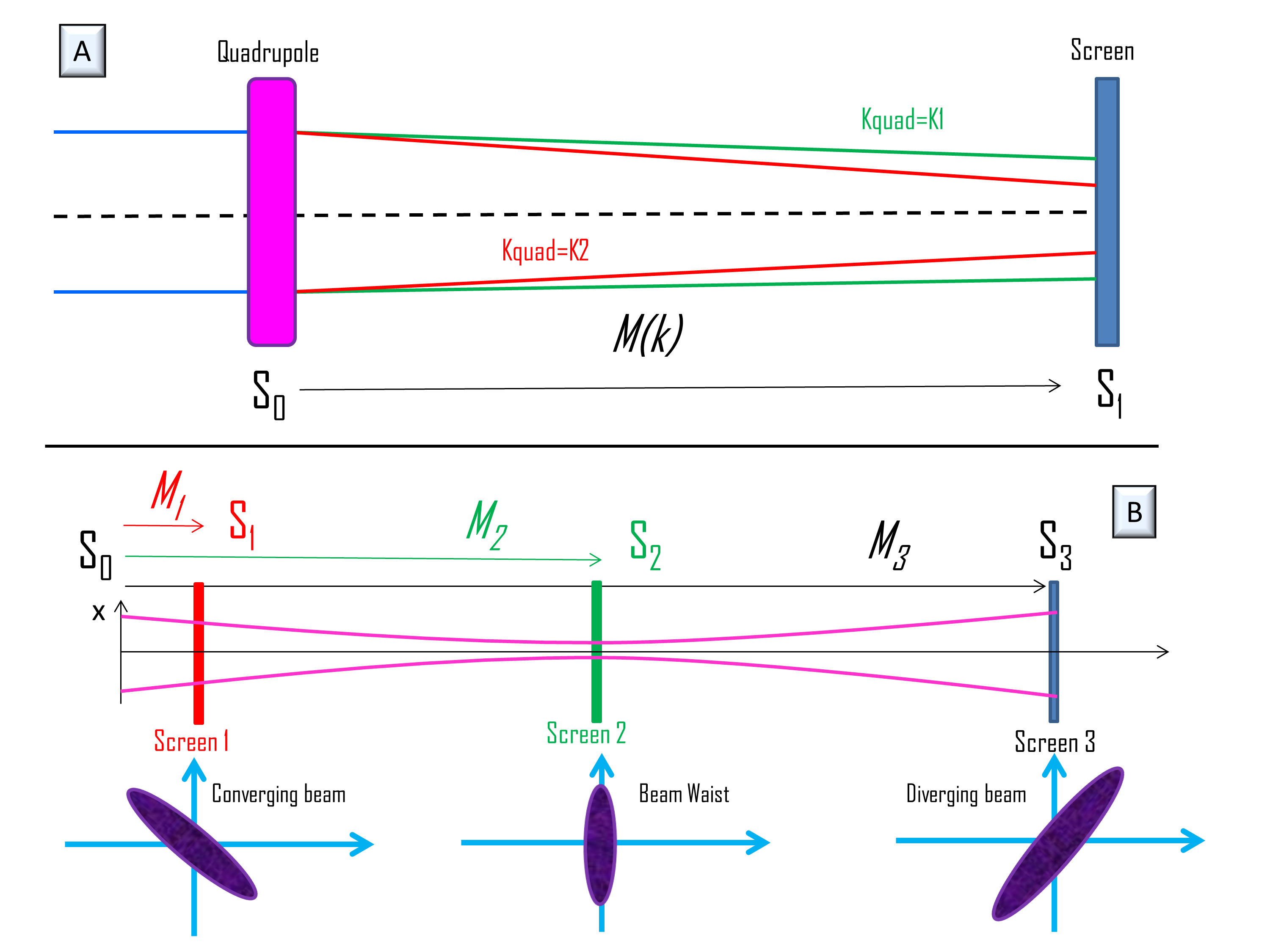}
\caption{Sketch representing the working principle of the quadrupole scan (A) and multi-screen (B) for measuring the beam emittance. }
\label{quadscan}
\end{center}
\end{figure}

\subsubsection{Quadrupole Scan}
A sketch of the setup is depicted in Fig. \ref{quadscan} A. In the quadrupole scan measurements it is possible to retrieve the second order moments of the beam at the entrance of the first (or only) quadrupole used for the scan (position $s_0$). The lattice is composed by one or more quadrupoles followed by a drift space. At the end of the drift is placed the measurement screen (position $s_1$). The linear matrix used to describe the transport of the beam through the lattice is a function of the variable strength k of the quadrupole (or of the strengths $k_i$ for each quadrupole). From Eq. \ref{sigma1} and \ref{sigma2} we can extrapolate the relations:
\begin{equation} \label{qscan}
\begin{aligned}
& \sigma_{11,s1}(k_1)=M_{11}^2(k_1)\sigma_{11,s0}+2M_{11}(k_1)M_{12}(k_1)\sigma_{12,s0}+M_{12}^2(k_1)\sigma_{22,s0} \\
& \sigma_{11,s1}(k_2)=M_{11}^2(k_2)\sigma_{11,s0}+2M_{11}(k_2)M_{12}(k_2)\sigma_{12,s0}+M_{12}^2(k_2)\sigma_{22,s0} \\
& ... \\
& \sigma_{11,s1}(k_n)=M_{11}^2(k_n)\sigma_{11,s0}+2M_{11}(k_n)M_{12}(k_n)\sigma_{12,s0}+M_{12}^2(k_n)\sigma_{22,s0}
\end{aligned}
\end{equation}
where $n$ is the number of $k$ values which are scanned during the measurement. Since we need to measure three unknown parameters, i.e. $\sigma_{11,s0}$, $\sigma_{12,s0}$ and $\sigma_{22,s0}$, we need to collect at least three measurement of the beam-size at position $s_1$ for different values of $k$, which correspond to different currents of the quadrupole used for the scan \cite{cern}.\\
The quadrupole scan measurement can be perturbed by several effects. First of all the space charge effect must be either included in the transport matrices used to analyze the data or must be negligible. Since the quadrupole scan is typically used at high energy, the space charge is usually not included. There are analogous techniques anyway, such as the solenoid scan which might be used in the injector region and therefore for lower beam energies. In this case the implementation of the space charge in the model plays an important role \cite{emittanceregae}.\\
Another limitation of the quadrupole scan is constituted by the chromatic effects produced by a high energy spread of the beam under analysis. Those effects can produce unwanted emittance dilution that perturbs the results, as described in \cite{energyspread}. A small beamsize inside the quadrupole halps with damping this error, nevertheless achieving this condition when measuring beams from plasma is technically difficult because the quadrupole used for the scan needs to be placed very close to the exit of the plasma target (e.g. closer than 50cm). 

\subsubsection{Single-shot measurement with permanent quadrupoles and energy-spectrometer}
The setup of this measurement is constituted by a set of permanent quadrupoles followed by a dipole-spectrometer and a screen. The sketch can be found for example in reference \cite{weingartner}. This technique is a single shot, nevertheless a 'scan' internal to the bunch itself is performed. Specifically the spot-size of the electrons with deviations from the reference energy is analyzed. The method implicitly assumes that the local emittance along slices in the input beam having slightly different energy is constant. A single-shot image is recorded on a screen downstream the dipole spectrometer. For the data analysis Eq. \ref{sigma2} combined with the definition of the quadrupole strength 
\begin{equation}
k[m^{-2}]=0.2998\frac{g[T/m]}{p[GeV/c]}
\end{equation}
can be used. In the equation above $g$ is the gradient of the quadrupole and $p$ is the beam-momentum \cite{rossbach}.
This technique is popular in the field of plasma accelerators \cite{weingartner} \cite{barber} \cite{li} because it is single-shot. The drawback of this method is that it is intrinsically limited to  high-energy spread beams, indeed in the limit of a mono-energetic beam there would be no internal spread of $k$ which could be analyzed. 

\subsubsection{Multi-Screen Method}
The multi-screen method is equivalent to the conventional quadrupole scan technique but, instead of varying the currents of one quadrupole, we collect images at different screen-locations while keeping the machine-optics fixed. A sketch of the setup is shown in Fig. \ref{quadscan} B. The optics of the lattice has to be chosen in such a way that one of the screens intercepts the beam waist and the other screens (minimum two) have sufficient phase-advance increment from the central one to resolve the focusing and defocusing of the trace-space-beam-ellipse. Typically this is realized by inserting quadrupoles between the screens. Also in this case we analyze the data by using Eq. \ref{sigma2} but the $M$ matrix will be different for each screen. The main advantage of the multi-screen method w.r.t. the quadrupole scan is that the beam optics is not fixed. The measurement is not single-shot but can be done parasitically to beam operation using kickers \cite{gerth}.\\
It is currently being investigated if this measurement could be performed single-shot by implementing non-destructive OTR screens \cite{delerue2} \cite{thomas2}. This possibility is of course especially attractive for the plasma-community, nevertheless this method would be usable only for high beam energies when the scattering at the screen-locations is negligible. 

\subsubsection{Reconstruction of the 4D-Phase Space}
In the paragraph above we have always implicitly assumed that the motion in x and y of our beam can be decoupled and we have looked at the separate behaviors of the emittances in the vertical and horizontal planes. If this assumption is not valid we need to calculate also the correlation terms between x and y \cite{barletta}. Most of the methods that we have described so-far can be extended to retrieve the 4D-transverse-emittance of the beam, as for example experimentally shown in \cite{TEM_4D}, \cite{quadscan_4D} \cite{hock}.

\section{Overview of Techniques for Longitudinal Phase Space Characterization}
Longitudinal phase space characterization consists on the measurements of the energy profile of the beam, the longitudinal current profile and the correlation between the two. While many techniques presently allow for the characterization of the bunch length, we will see that the characterization of the local beam properties along longitudinal slices of the bunch is mostly done using Transverse Deflection Structures (TDSs).  
\subsection{Energy and Energy Spread Measurements}
The momentum of a charged particle can be expressed in terms of its orbit radius inside a magnetic field, the strength of the named magnetic field, and the charge of the particle \cite{conte}. For a particle with the same charge as the electron, it is useful to remember:
\begin{equation}
p_c \left[ \frac{GeV}{c} \right] = 0.2998 \frac{B_0\left[T\right]}{\rho \left[ m \right]}
\end{equation}
By measuring the magnetic field $B_0$ of a dipole, bending the electron bunch along a known trajectory with curvature radius $\rho$, it is possible to calculate the central momentum of the beam $p_c$. Similarly by measuring the bunch width at the exit of the dipole it is possible to calculate the energy spread of the beam. For this calculation the Dispersion Function $D_x (z)$ at the $z$ location of the measurement screen is needed. The Dispersion Function is defined such that $\frac{\Delta E}{E_c} \cdot D_x$ determines the offset of the reference trajectory from the ideal path for particles with a energy spread deviation $\frac{\Delta E}{E_c}$ from the ideal momentum $p_c=\frac{E_c}{c}$ (for a relativistic beam). The momentum of the off-energy electrons can be calculated using the equation:
\begin{equation}
p_d=p_c \left( 1+\frac{\Delta x}{D_x} \right)
\end{equation}
Since the beam size at the measurement screen is not only determined by the dispersive contribution but also by the betatron oscillations of the beam in the transverse planes such that $\sigma_{x,meas}=\sqrt{\epsilon_x \beta_x + D_x^2 \sigma_{\delta}^2}$, the profile of the energy distribution can be retrieved with a resolution $\sigma_{E,res}$ defined as \cite{wiedemann}:
\begin{equation}
\sigma_{E,res}=2 \frac{\sqrt{\epsilon_x \beta_x}}{D_x}\left\langle E \right\rangle .
\end{equation} 
The resolution of the measurement can be optimized by properly choosing the bending angle of the spectrometer (affects $D_x$), the beam optics before the spectrometer (can be used for matching $\beta_x$ at the screen) and possible additional magnetic elements between the dipole spectrometer and the measurement screen (can be used for matching both $D_x$ and $\beta_x$).

\subsection{Bunch Length and Characterization of the longitudinal Slices of the Beam}
The characterization of the longitudinal properties of the beam does not cover exclusively the measurement of the bunch length but also includes for example the reconstruction of the longitudinal current-profile of the beam as well as longitudinal-slice-properties of the electrons such as slice energy spread and slice emittance.\\
We will divide the techniques described in this paragraph into two groups:
\begin{itemize}
\item Direct particle measurement techniques;
\item Radiative techniques.
\end{itemize} 

\begin{figure}[ht]
\begin{center}
\includegraphics[width=14cm]{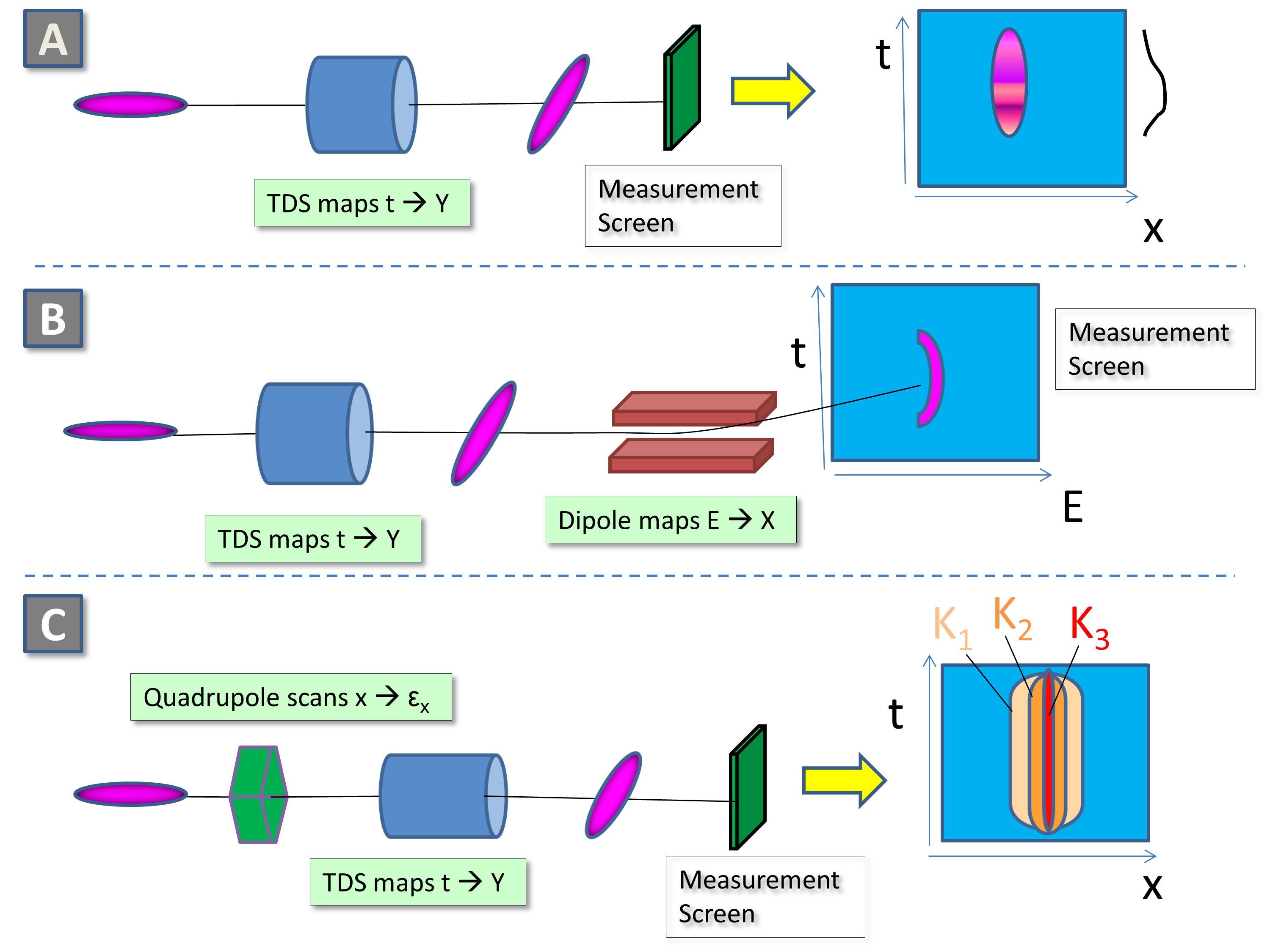}
\caption{Sketch representing the working principle of slice-characterization measurements using a Transverse Deflection Structure: current profile (A), longitudinal phase-space (B) and slice-emittance (C). }
\label{tds}
\end{center}
\end{figure}

\subsubsection{Direct Particle Measurement}
Direct Particle Measurement techniques detect directly the particles of the electron bunch. In order to get information concerning the longitudinal profile of the beam, that cannot be directly detected, we need to introduce a correlation between the longitudinal distribution of the beam and a transverse property of the beam that can be measured on a screen.\\
A relatively old technique is for example the so-called \emph{longitudinal tomography} which uses a RF-booster in combination with an energy spectrometer. This technique is multi-shot. The beam travels in a RF-booster where a linear energy chirp, i.e. a linear correlation between electron position relative to the bunch center and energy, is introduced. Downstream the booster is placed an energy spectrometer allowing for the measurement of the energy-spread distribution for different RF-phases of the booster. By analyzing the measured energy profiles with a tomographic algorithm it is possible to reconstruct the longitudinal phase space of the beam\cite{malyutin}. The time-resolution of the measurement depends on the energy chirp impressed to the beam in the RF-booster (high RF-frequencies help, see for example new developments using THz cavities \cite{vinatier}) and the dispersion in the spectrometer (high dispersion helps). Moreover the initial energy spread of the beam must be small enough and the space charge effect must be negligible.\\
Compared to longitudinal beam tomography methods, \emph{Transverse Deflection Structures} (TDS) offer simpler measurements and better longitudinal resolution. That is why they have rapidly spread out in most of the particle accelerators.\\
The working principle of the TDS is the following: the longitudinal distribution of the e-bunch is mapped into the transverse one thanks to the time dependent transversely deflecting field \cite{LOLA}, for example directed in the vertical direction $y$. A measurement screen downstream the structure is used for measuring the transverse size of the beam in the streaking direction of the structure, assumed to be the vertical in the example shown in Fig. \ref{tds} A. The resolution $R_t$ of the measurement is qualitatively given by the ratio between the vertical betatron size of the beam at the screen when the TDS is switched off and the one when the TDS is switched on, the latter quantity is also often called shearing parameter $S_{y,t}$. The resolution is given by \cite{emma, roehrs}:
\begin{equation}
R_t=\frac{\sigma_{y,off}}{S_{y,t}}=\frac{\sqrt{\epsilon_y \beta_y(s)}}{\sqrt{\beta_y (s) \beta_y (s_0)} sin(\Delta \phi_y)} \frac{E}{\omega e V_0}.
\end{equation}  
In the Equation above: $E$ is the beam kinetic energy, $\epsilon_y$ is the vertical geometric emittance, $\Delta \phi_y$ is the beam-phase-advance in the vertical plane, $\beta_y$ is the Courant Snyder beam beta function, $f=\frac{\omega}{2 \pi}$ is the RF frequency of the structure and $V_y$ is its peak deflection voltage. Moreover we have called $s_0$ the central longitudinal position of the structure in the beamline and $s$ the position of the measurement screen. \\
Transverse Deflection Structures have proved to be capable to deliver sub-fs longitudinal resolution using high frequency (X-band) technology \cite{behrens} \cite{pietro}. The ultimate limitation for going below those values is represented by the parasitic perturbation of the energy distribution of the beam in the TDS which is a direct consequence of the Panofsky-Wenzel-Theorem \cite{pwt}. It can be demonstrated that the uncorrelated energy spread increase in the TDS is inversely proportional to the resolution of the measurement.\\
Despite this limitation, TDS allow a comprehensive characterization of the electron beam including:
\begin{itemize}
\item measurement of the longitudinal profile of the beam without any assumption regarding the shape of the beam distribution (see Fig. \ref{tds} A);
\item combining the TDS with a spectrometer dipole, the latter streaking the beam in the perpendicular direction with respect to the streaking imparted by the TDS, it is possible to measure the longitudinal phase space of the beam (see Fig. \ref{tds} B) \cite{longphasespace}. The energy spread induced by the TDS ultimately limits the characterization of the longitudinal phase space of the beam for high resolution measurements. 
\item combining the TDS with a quadrupole scan or with a multi-screen-lattice, it is possible to measure the slice emittance of the beam on the plane perpendicular to the streaking direction of the TDS (see Fig. \ref{tds} C). If the projections of the beam at the screen are analyzed using a tomographic algorithm, it is possible to retrieve the transverse-trace-space distribution of each slice \cite{sliceemittance}.
\end{itemize}
Besides those well-known measurements, there are also many new developments ongoing involving the extension of this concept. The development of TDS cavities capable of streaking along a variable direction is currently under-testing \cite{polarix}. This novel device would open new opportunities for a broader characterization of the particle distribution including for example the reconstruction of the 3D charge density of the beam \cite{daniel_tomo}.\\
Another branch of research involves experiments on alternative technologies to produce high frequency TDS. Technology under consideration are for example plasmas \cite{irene} and THz structures \cite{THzTDS1}, \cite{THzTDS2}.

\subsubsection{Radiative Techniques}
Radiative techniques relies on the analysis of radiation which is generated by the particle beam under analysis and encodes information concerning its bunch length. Those methods require in general a setup which is much more compact than the techniques described in the previous chapter. Moreover they can be non-destructive. For those reasons radiative techniques are often used for tuning the working point of the accelerator and for feedback systems. Their drawback is that the analysis of the data to reconstruct the current profile of the beam is more complex than the direct particle detection techniques.\\
Radiative techniques include many types of measurements. Due to space restrictions, in this lecture we will cover only the measurements based on coherent radiation generation and the electro-optic sampling (EOS) technique.\\
Coherent radiation measurements are based on the fact that the coherent part of the spectrum of the radiation emitted by charged particles is linked to the three-dimensional charge distribution of the particles in the bunch. We define the three dimensional Form Factor function $F(\lambda, \Omega)$ according to\cite{formfactor}:
\begin{equation} \label{ff1}
\frac{d^2 U_{b}}{d \lambda d \Omega}= \left(\frac{d^2 U_{sp}}{d \lambda d \Omega}\right) \left[ N+N(N-1) \mid F(\lambda, \Omega) \mid ^2 \right] 
\end{equation} 
\begin{equation} \label{ff2}
F(\lambda, \Omega)=\int S_{3D}(\overline{r})\exp{(-i\overline{k\cdot \overline{r}})} d\overline{r}
\end{equation} 
where $\frac{d^2 U_{b}}{d \lambda d \Omega}$ is the spectrum of the radiation emitted by the bunch, $\frac{d^2 U_{sp}}{d \lambda d \Omega}$ is the spectrum of the radiation emitted by a single particle, $N$ is the number of particles in the bunch and $S_{3D}$ is the 3D normalized particle density distribution. The coherent part of the spectrum is represented by the second additive term in Eq. \ref{ff1} and it is therefore proportional to $N^2$. If longitudinal and transverse distributions are uncorrelated and under the assumption that the radiation is confined in a narrow cone in the forward direction and therefore the observation angle is small, it is possible to replace the 3D Form Factor Function with the 1D longitudinal Form Factor function.  
Coherent radiation measurement techniques rely on a standard \emph{recipe} for measuring the bunch length and current distribution:
\begin{itemize}
\item First of all, conditions such that the particle beam emits coherent radiation must be realized. For this purpose many types of radiation can be used, such as Transition and Diffraction Radiation \cite{transdiffractionradiation}, Synchrotron Radiation, Smith-Purcell Radiation \cite{smithpurcell} etc. Some of the radiation production methods are destructive (e.g. Transition Radiation), while others might be non-destructive (e.g. Diffraction Radiation or Synchrotron Radiation). Each radiation type will have a characteristic single particle spectrum $\frac{d^2 U_{sp}}{d \lambda d \Omega}$ in Eq. \ref{ff1}.
\item Then, the spectrum of the emitted radiation has to be measured. This step can happen in a single-shot (e.g. using gratings-based spectrometers) or by multi-shot (e.g. if interferogram-based spectrometers are used, such as Martin-Puplett spectrometer). In order to retrieve the spectrum of the radiation emitted by the particle-beam, it is critical to correct the measured spectrum at the detector to account for the imperfections of the optical-transport-line (e.g. transmission of vacuum windows, detector response etc.).
\item Finally, the longitudinal distribution of the beam can be retrieved from the computed radiation spectrum at the source location. This can be done by calculating directly the amplitude of the Form Factor function and applying a phase retrieval method (such as e.g. Kramers-Kronig relation \cite{kk}) for calculating its phase. It is important to stress that at this point some assumptions concerning the longitudinal shape of the beam distribution have to be made. For example the Kramers-Kronig relation retrieves only the minimum phase of the Form Factor, e.g. it computes the shape with the least complex substructure compatible with the radiation spectrum.
\end{itemize}
In order to get the correct reconstruction of the longitudinal shape of the beam profile, the range of the measured spectrum has to be as large as possible (it has to cover all the main features of the bunch profile) \cite{ctrwidespectrum}.
Complex single-shot CTR spectrometers are in use in many facilities \cite{ctrdresden}, \cite{ctrhighres}, \cite{ctrhighres2},\cite{ctrslac}. Recently sub-fs resolution has been obtained using this method applied to beams produced by plasmas \cite{ctrhighres3}.\\
Electro-Optic Sampling (EOS) is another popular method which is often employed for non-destructive single-shot bunch diagnostics. EOS is based on the conversion of the Coulomb field of the particle bunch into a variation of optical intensity. In order to achieve this, the bunch passes close to an electro-optical crystal (tipically ZnTe or GaP). The Coulomb field of the bunch induces a change in the refractive index of the crystal (Pockels effect). The information about the longitudinal profile is encoded in a refractive index change which can be converted into an intensity variation by means of a laser together with polarizers. There are several possibilities for retrieving the information about the longitudinal bunch profile from the analysis of the detected pulse. The Spectral Decoding technique is the simplest method \cite{spectraldecoding}, but also Spatial Encoding \cite{spatialencoding}, and Temporal Decoding \cite{temporaldecoding} methods have been experimented. The latter so far delivered the best temporal resolution, corresponding to few tens of fs. EOS is not only used for online monitoring of the bunch length but also as arrival time monitor.

\end{document}